\documentclass[12pt,preprint]{aastex}
\usepackage{psfig,epsfig,amsfonts,amsmath,amssymb,graphicx,morefloats,appendix,tensor}
\usepackage{color}
\usepackage{natbib}
\usepackage{hyperref}
\usepackage{subfigure}
\begin{document}

\slugcomment{Accepted: September 9, 2022}

\title{The Mouse that Squeaked: A small flare from Proxima Cen observed in the millimeter, optical, and soft X-ray with \textit{Chandra} and ALMA}

\author{Ward S. Howard\altaffilmark{1}, Meredith A. MacGregor\altaffilmark{1}, Rachel Osten\altaffilmark{2,3}, Jan Forbrich\altaffilmark{4,5}, Steven R. Cranmer\altaffilmark{1,6}, Isaiah Tristan\altaffilmark{1}, Alycia J. Weinberger\altaffilmark{7}, Allison Youngblood\altaffilmark{6,8}, Thomas Barclay\altaffilmark{8,9}, R. O. Parke Loyd\altaffilmark{10}, Evgenya L. Shkolnik\altaffilmark{10}, Andrew Zic\altaffilmark{11,12}, David J. Wilner\altaffilmark{4}}

\altaffiltext{1}{Department of Astrophysical and Planetary Sciences, University of Colorado, 2000 Colorado Avenue, Boulder, CO 80309, USA}
\altaffiltext{2}{Space Telescope Science Institute, Baltimore, MD 21218 USA}
\altaffiltext{3}{Center for Astrophysical Sciences, Johns Hopkins University, Baltimore, MD 21218, USA}
\altaffiltext{4}{Center for Astrophysics \textbar~Harvard \& Smithsonian, Cambridge, MA 02138, USA}
\altaffiltext{5}{Centre for Astrophysics Research, University of Hertfordshire, AL10 9AB, UK}
\altaffiltext{6}{Laboratory for Atmospheric and Space Physics, University of Colorado, Boulder, CO 80303, USA}
\altaffiltext{7}{Earth \& Planets Laboratory, Carnegie Institution for Science, Washington, DC 20015, USA}
\altaffiltext{8}{NASA Goddard Space Flight Center, Greenbelt, MD 20771, USA}
\altaffiltext{9}{University of Maryland, Baltimore County, Baltimore, MD 21250, USA}
\altaffiltext{10}{School of Earth and Space Exploration, Arizona State University, Tempe, AZ 85287, USA}
\altaffiltext{11}{School of Mathematical and Physical Sciences, and Research Centre in Astronomy, Astrophysics and Astrophotonics, Macquarie University, NSW 2109, Australia}
\altaffiltext{12}{Australia Telescope National Facility, CSIRO, Space and Astronomy, PO Box 76, Epping, NSW 1710, Australia}

\begin{abstract}
We present millimeter, optical, and soft X-ray observations of a stellar flare with an energy squarely in the regime of typical X1 solar flares. The flare was observed from Proxima Cen on 2019 May 6 as part of a larger multi-wavelength flare monitoring campaign and was captured by \textit{Chandra}, LCOGT, du Pont, and ALMA. Millimeter emission appears to be a common occurrence in small stellar flares that had gone undetected until recently, making it difficult to interpret these events within the current multi-wavelength picture of the flaring process. The May 6 event is the smallest stellar millimeter flare detected to date. We compare the relationship between the soft X-ray and millimeter emission to that observed in solar flares. The X-ray and optical flare energies of 10$^{30.3\pm0.2}$ and 10$^{28.9\pm0.1}$ erg, respectively, the coronal temperature of $T$=11.0$\pm$2.1 MK, and the emission measure of 9.5$\pm$2.2$\times$10$^{49}$ cm$^{-3}$ are consistent with M-X class solar flares. We find the soft X-ray and millimeter emission during quiescence are consistent with the G\"udel-Benz Relation, but not during the flare. The millimeter luminosity is $>$100$\times$ higher than that of an equivalent X1 solar flare and lasts only seconds instead of minutes as seen for solar flares.
\end{abstract}


\section{Introduction}\label{sec:intro}
The vast majority of terrestrial planets suitable for atmospheric characterization with the James Webb Space Telescope (JWST) and extremely large telescopes orbit nearby M-dwarfs \citep{Kempton:2018, DiamondLowe:2020}. M-dwarfs are known to flare regularly throughout their lifetimes \citep{Mohanty_Basri:2003, Houdebine:2003, Tarter:2007, France2020, Loyd:2021}, driving the composition and even survival of terrestrial atmospheres \citep{Segura:2010, Tilley:2019, Chen:2021}. Stellar flares emit radiation across the electromagnetic spectrum as a result of particle acceleration and plasma heating following magnetic reconnection in the stellar atmosphere \citep{Kowalski:2013}. Particles of different  energies brake at different depths in the stellar atmosphere and produce emission at different wavelengths \citep{Klein_Dalla:2017}. Simultaneous multi-wavelength observations are needed to better understand the processes at work throughout a flaring event because different wavelengths probe different components of the flare structure and evolution as well as different physical processes and parts of the stellar atmosphere \citep{MacGregor:2021}.

While the multi-wavelength properties of moderate-to-large stellar flares ($\geq$10$^{31}$ erg) have received a large amount of recent attention (e.g. \citealt{Kowalski2019, Namekata:2020, Howard:2020, MacGregor:2021}), the multi-wavelength properties of small flares have not received the same degree of attention (e.g. \citealt{Kowalski:2016, Paudel:2021, Zic:2020}) despite their high frequency and connection to space weather. In the heliophysical context, space weather consists primarily of accelerated particles and coronal mass ejections (CME); in the M-dwarf context X-ray and UV emission from flares also become significant components of space weather \citep{Loyd:2018b}. Within the solar system, even moderate space weather events can induce significant disequilibrium states in the Martian atmosphere \citep{Kajdic:2021}. For example, solar energetic particles associated with an X9-class solar flare on Dec 5, 2006 induced an order of magnitude increase in the atmospheric escape rate of Mars \citep{Futaana:2008}. For terrestrial planets in close orbits around M-dwarfs, smaller M and X class flares and associated particle emission would likely have similar or greater effects than an X9 flare from the Sun. Small flares from M-dwarfs provide a unique opportunity for multi-wavelength comparisons between solar and stellar contexts because the energy range of these events is most similar to the energy range seen during large solar flares. Flares from mid-to-late M-dwarfs remain detectable down to very low energies (e.g. energies of 10$^{27}$ erg in $U$ \citep{Lacy:1976, Walker:1981}), typical of flares routinely observed from the Sun. Solar flares are often recorded with comprehensive multi-wavelength coverage and spatial resolution, enabling insights into the physical mechanisms responsible for flare emission at all wavelengths, including magnetic reconnection, particle acceleration, and the resulting heating of the plasma \citep{Benz:2017}.

Recent monitoring of M-dwarf flare stars with the Atacama Large Millimeter Array (ALMA) has revealed the unexpected presence of millimeter flares, apparently common events that had gone largely undetected until now (e.g. \citealt{MacGregor:2018, MacGregor:2020, MacGregor:2021}). Millimeter flaring was detected by \citet{MacGregor:2018} in ALMA data obtained during a search for debris disk emission around Proxima Cen \citep{Anglada:2017}, leading to dedicated searches for more flare events. Stellar flares at radio frequencies of 10-20 GHz have been previously detected \citep{Gudel:2002}, although the spectral energy distribution of stellar flares from radio to millimeter frequencies is not yet clear \citep{MacGregor:2020, MacGregor:2021}. Millimeter flares from the Sun occur on timescales of minutes and trace particle acceleration in flare loops, with $<$X6 flares exhibiting a falling spectral index with frequency, typical of gyrosynchrotron emission \citep{Krucker:2013}. Millimeter flares corresponding to $\geq$X6 events often display a steeply positive spectral index suggestive of interactions with relativistic particles produced in nuclear processes \citep{Krucker:2013, Wedemeyer:2016}. The millimeter luminosity of the solar events generally correlates with the 1-8 $\AA$ GOES soft X-ray emission. On the other hand, millimeter flares from M-dwarfs have characteristic timescales of seconds, have $\geq$10$\times$ higher luminosities than solar flares, and steeply negative spectral indices \citep{MacGregor:2020, MacGregor:2021}. In addition to ALMA, several millimeter flares have been observed with high luminosities by the Atacama Cosmology Telescope and the South Pole Telescope 3G \citep{Naess:2021, Guns:2021}. None of these were observed simultaneously in the X-ray, although several had optical counterparts. 

An extreme flare from a young stellar object (YSO) was simultaneously captured by \textit{Chandra} and the Berkely-Illinois-Maryland Association (BIMA) array \citep{Bower:2003}, but this event may not resemble the much smaller stellar flares typical of main sequence M-dwarfs \citep{Getman:2021}. The \citet{Bower:2003} flare reached a peak soft X-ray luminosity of nearly 10$^{33}$ erg s$^{-1}$, three orders of magnitude higher than an X1 analogue flare. Flares from main sequence stars have never been observed simultaneously with millimeter and soft X-ray (SXR) data, making it difficult to place them in the solar context.

Here, we present the first main sequence stellar flare with simultaneous millimeter and SXR observations from ALMA and \textit{Chandra}, alongside optical photometry and spectroscopy. This flare is only the second millimeter event reported with broad multi-wavelength coverage. The flare was observed on 2019 May 6 as part of a larger $\sim$40 hr multi-wavelength monitoring campaign of Proxima Cen. Proxima Cen is an M5.5 dwarf at a distance of 1.3 pc and host to a small temperate planet \citep{AngladaEscude:2016}, making it an ideal proxy for the host stars of most JWST terrestrial planet targets. Proxima Cen rotates with a period of 83 d and remains flare-active \citep{Benedict:1998, Vida:2019}. The May 6 event was also captured in the optical by the du Pont telescope at Las Campanas Observatory and the Las Cumbres Observatory Global Telescope (LCOGT). The flare was not observed by the other observatories in the campaign: the Neil Gehrels Swift Observatory, Hubble Space Telescope (HST), Evryscope, the Transiting Exoplanet Survey Satellite (TESS), and the Australian Square Kilometre Array Pathfinder (ASKAP) as they were not observing at the time.

\begin{figure*}
	\centering
	{
		\includegraphics[width=0.98\textwidth]{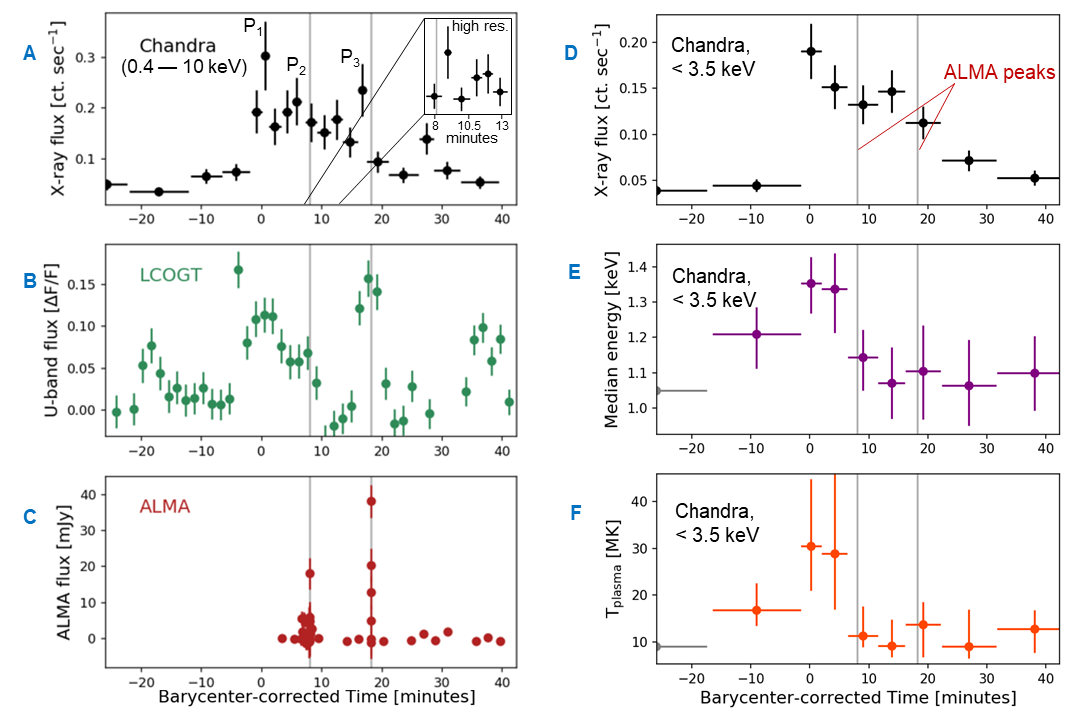}
	}
	\caption{X-ray to millimeter light curves of the flare aligned to the barycentric JD \textit{Chandra} peak time of 2458609.751 (TDB). The \textit{Chandra} light curve with adaptive binning is shown in Panel A. An inset image at a fixed temporal resolution of 1 min in Panel A shows a possible increase in X-ray flux near the time of the first ALMA peak. A large increase during the X-ray decay phase correlates well with both LCOGT $U$-band and ALMA flux increases in Panels B and C. On the right, a binned X-ray light curve (Panel D) and the median energy (E) and plasma temperature (F) are shown. Grey lines are times of ALMA peaks.}
	\label{fig:chandra_overview}
\end{figure*}

\begin{figure*}
	\centering
	\subfigure
	{
		\includegraphics[width=0.98\textwidth]{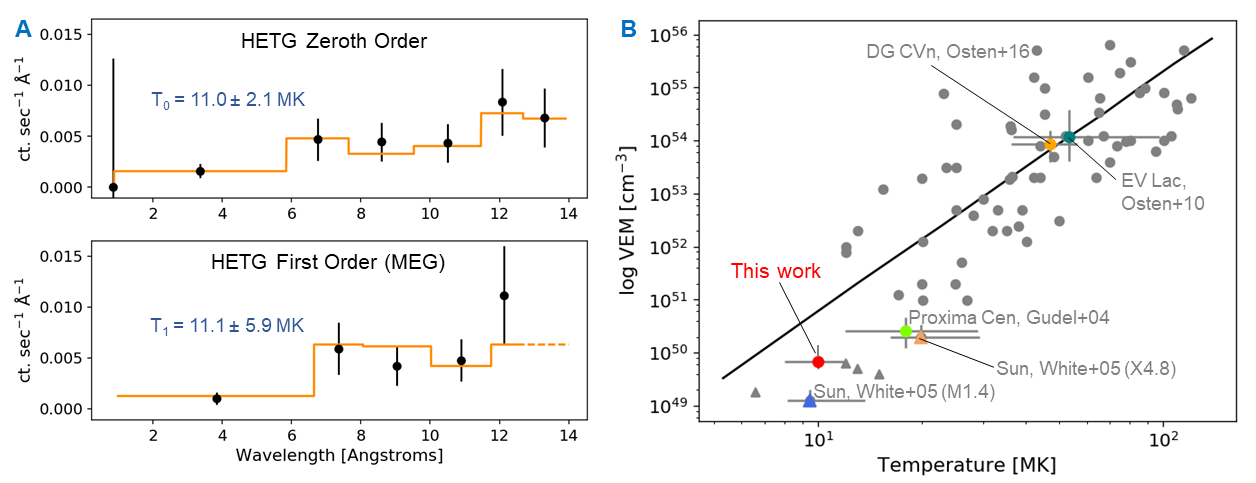}
	}
	\subfigure
	{
		\includegraphics[width=0.98\textwidth]{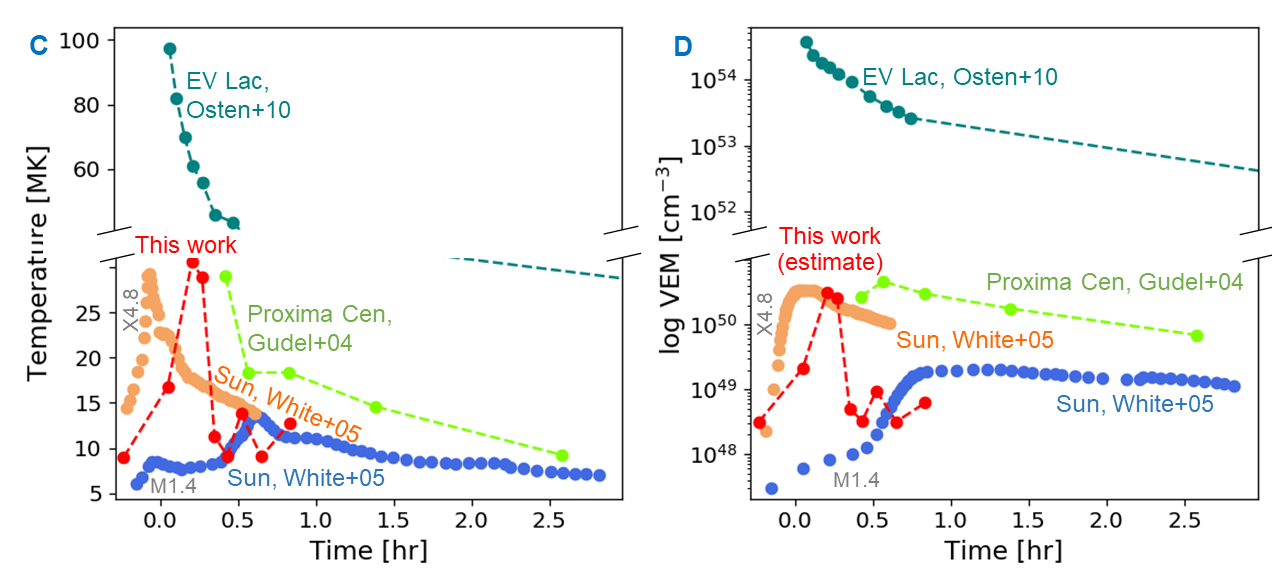}
	}
	\vspace{-0.2cm}
	\caption{While stellar flares can reach high temperatures and emission measures (EM), \textit{Chandra} HETG spectra place the May 6 event squarely in the temperature-EM regime of M/X-class solar flares. Panel A: Zeroth and first order HETG flare spectra (black) are fit with an APEC temperature model (orange). The integrated temperature is close to the 12.8 MK median of the MASME temperatures in Fig. \ref{fig:chandra_overview}. Panel B: The best-fit integrated temperature and EM are compared with a range of solar and stellar flares \citep{Getman:2021, Osten:2010, Osten:2016}. Dots are stellar flares, triangles are solar flares. Highlighted flares are color-coded, with grey flares being the \citep{Getman:2021} main-sequence stellar flare sample. Our X1-analogue is comparable to two M1.4 and X4.8 class solar flares from \citet{White:2005}. The \citet{Getman:2021} stellar $T$-EM fit is shown and used to estimate the EM of our flare. Panels C \& D: The time evolution of the flare's estimated temperature and potential EM are compared with a stellar superflare \citep{Osten:2010}, a typical stellar flare \citep{Gudel:2004b}, and two solar flares from \citet{White:2005}.}
	\label{fig:hetg_orders}
\end{figure*}

\section{Multi-wavelength Flare Observations}\label{multiwave_data}
Multi-wavelength observations and reduction details for each observatory are described below. Because multi-wavelength flare campaigns require a large amount of data reduction from very different instruments, we break up the data reduction by observatory and wavelength to aid the reader in locating details for each wavelength.

\subsection{X-ray Observations with \textit{Chandra}}\label{sec:chandracxc_obs}
Coordinated \textit{Chandra} observations of the flare were taken with the ACIS-S detector in the HETG grating configuration under a Cycle 20 DDT program (PI: MacGregor; observation IDs: 22185 and 22186). \textit{Chandra} observed the target in the faint timed exposure mode for a combined 8.3 hr on 2019 May 3 and 2019 May 6 and recorded only the one event on 2019 May 6. To determine this, we produced both uniform and adaptively-binned light curves of the other times of observation and did not observe any increase in the count rate comparable to the flare. The quiescent emission outside the flare was also detected, with minor variability present in the light curve. The flare began at 5:51 UT and lasted 38 min. A light curve is constructed from all zeroth and first order events recorded in the HETG Level 2 events file. We adaptively bin the counts in units of counts s$^{-1}$ to maximize the features and apply a barycentric correction in panel A of Figure \ref{fig:chandra_overview}. The variable time per bin is chosen to include 20 events per bin, as this number provides a balance between S/N and time-resolution. Uncertainties are given by Poisson statistics for the number of counts in each bin.

The flare is complex, with a rapid rise followed by a more gradual peak and then a third impulsive event during the decay phase. The more gradual middle peak appears strongest at low ($<$2 keV) energies. Due to the small size of the flare, the complex peaks are separated from each other at low significance as shown in Figure \ref{fig:chandra_overview}.

The event emitted a peak flux of 10$^{27.3\pm0.2}$ erg s$^{-1}$ in the HETG bandpass ($\sim$2-30$\AA$) and 10$^{30.3\pm0.2}$ erg integrated over the entire flare duration of $\sim$40 min. The count rates from the light curve are converted into energies separately for the zeroth and first orders using PIMMS version 4.11 \footnote{See https://cxc.harvard.edu/toolkit/pimms.jsp}. For the conversion, we assume a 10 MK plasma/APEC model, 0.4 solar abundance, and neutral hydrogen column density of 3$\times$10$^{18}$ cm$^{-2}$ for the flare. Before converting to energies, we subtract the zeroth and first order background count rates of 0.023 and 0.035 ct s$^{-1}$, respectively. The flux of each light curve bin is multiplied by the bin width in seconds and scaled for the distance of Proxima Cen to obtain the energies in each bin. Summing the energy bins produces a zeroth order energy log $E_0$=30.35$\pm$0.2 and first order energy log $E_1$=30.24$\pm$0.2 erg. We perform a weighted average of $E_0$ and $E_1$, weighting by the fraction of counts in each order to obtain a final energy of log $E_X$=30.3$\pm$0.2 erg. Converted to the GOES 1-8 $\AA$ band in PIMMS, we find the flare reached 10$^{26.5\pm0.1}$ erg s$^{-1}$ at peak. We measure the peak flux in the GOES 1-8 $\AA$ bandpass at a distance of 1 au to be 10$^{-4}$ W m$^{-2}$, equivalent to an X1 solar flare. At 0.05 au, the 400$\times$ higher flux of 0.04 W m$^{-2}$ would have a much greater impact than an X1 flare at 1 au.

The median energy per bin at a reduced time resolution of 50 events per bin is shown for the 0.3-3.5 keV range in panel E. This energy range was chosen after inspecting histograms of the energy distribution of the events in each bin. Events below 3.5 keV show approximately Poissonian distributions in each bin while random noise is present at $\sim$5-10 keV which would otherwise bias the median energies. Uncertainties in median energy are obtained by bootstrapping the events in each time bin with replacement and recomputing the median across 10,000 MC trials. Due to the low number of counts, we use the geometric mean approximation to the median. In panel F, we estimate the temperature in MK using a scaling relation produced with the method of adaptively smoothed median energy (MASME) from \citet{Getman:2008}. 

\citet{Getman:2008} simulated a grid of flare counts, median energies, and temperatures and subsequently propagated the flares through the \textit{Chandra} ACIS-I instrument responses for various column densities. We selected the non-absorbed ACIS-I median energy to plasma temperature relation from Figure 1 of \citet{Getman:2008}. Since their model assumes the median energy was observed with ACIS-I while the median energies in our work are from ACIS-S, we ensure the difference in observed median energies is negligible. The largest difference in photon energy distributions should result from their slightly different effective areas. We therefore convolve a Gaussian for central positions at 1.4 keV and widths of 0.2 keV with the normalized effective areas of ACIS-S and ACIS-I, respectively. We then sample ACIS-I and ACIS-S photon energy distributions from the convolved functions and measure the difference in the observed median energies to be $\sim$0.01 keV. We also determine the scatter in the MASME median energy to temperature curve from Figure 1 of \citet{Getman:2008} to be $<$2 MK for energies below 1.4 keV. We therefore conclude the dominant source of uncertainty is the small number of photons in each time bin from which the median energies and error bars are computed, as is shown in panel E of Fig. \ref{fig:chandra_overview}. Further corroborating the MASME values, we find the mean temperature during the flare is qualitatively similar to the APEC temperature in the time-integrated spectra of panel A of Fig. \ref{fig:hetg_orders}.

We create zeroth and first order calibrated flare spectra (Fig. \ref{fig:hetg_orders}) using \texttt{CIAO} 4.13 as described below. The spectra are fit with an APEC model to determine the characteristic coronal temperature and emission measure (EM) during the flare. We create a new level 2 event file containing only the events during the flare and identify zeroth order source and background region files using \texttt{ds9}. We make a calibrated zeroth-order spectrum using the CIAO \texttt{specextract} script, which generates 0th order source and background pulse height analysis (PHA) files, response matrix files (RMFs), and auxiliary response files (ARFs). We use \texttt{tgextract} to make a cleaned first-order level 2 PHA file for the MEG and HEG arms, and the \texttt{mktgresp} script to create grating RMF and ARF files for each order and arm. Finally, we combine the spectra from the positive and negative arms of the HEG and MEG separately with the \texttt{combine\_grating\_spectra} script. The HEG spectrum lacked signal.

The zeroth and first order spectra are fit with a single temperature APEC model in XSPEC 12.12.0 \citep{Arnaud1996} to explore the plasma properties of the flare. We use SHERPA 4.14 to load the spectra, background, and response files and to subtract the background. We convert the spectra to wavelength and group the spectra by counts, with 7 count bin$^{-1}$. Wavelengths of $\geq$30 $\mathrm{\AA}$ are excluded by lack of counts. The abundance is set to $Z$=0.4 following \citet{Osten:2010} and frozen. We find similar effective temperatures $T$ and emission measures in separate fits to the zeroth and first order data as well as in a combined fit to all orders. Because the first order spectra are subject to higher uncertainties, we adopt the zeroth order values of $T$=11.0$\pm$2.1 MK and EM=9.5$\pm$2.2$\times$10$^{49}$ cm$^{-3}$. We do not fit multiple temperature components due to low counts in each wavelength bin. The resulting values of $T$=11.0$\pm$2.1 MK and EM=9.5$\pm$2.2$\times$10$^{49}$ cm$^{-3}$ are consistent with the plasma properties of both solar and stellar flares \citep{White:2005, Gudel:2004b}.

\subsection{Millimeter Observations with ALMA}\label{sec:alma_obs}
ALMA observed the star from 4:24 to 10:01 UT split across 4 scans of $\sim$1 hr each, recording two events as shown in Figure \ref{fig:chandra_overview}. ALMA data was taken with the Atacama Compact Array (ACA) using 9 antennas with baselines of 10 to 47 m. The observations were obtained on 2019 May 6 from 4:24 to 10:01 UT and split into 4 scheduling blocks (SB) of $\sim$1.5 hr each. Each SB was composed of 6.5 min scans (integrations) of the target interspersed with observations of the phase calibrator J1524-5903, resulting in 49 min on-source per SB. Flux and bandpass calibration were performed using the bright quasar J1517-2422 between each SB. The May 6 flare began in the X-ray and $U$-band during ALMA calibrations 5 minutes prior to the start of the second SB and continued for the first half hour of this block. 

To capture the smallest flares, the correlator was configured to maximize sensitivity to the continuum near 230 GHz. Spectral windows with a 2 GHz bandwidth each were observed with central frequencies of 225, 227, 239, and 241 GHz. The observations were carried out in dual polarization, enabling the XX and YY linear polarization to be measured. The ALMA pipeline was used to reduce the raw data, which relied on \texttt{CASA} version 5.1.1 \citep{McMullin:2007}. We then used the \texttt{clean} task in \texttt{CASA} to deconvolve and image the target. Light curves were produced by fitting point-source models directly to the visibilities to ensure accurate uncertainties. The one second minimum cadence of ALMA is used during the window around each flare seen in the X-ray, and 2 min or 10 sec cadence are used otherwise. No other sources in the field are sufficiently bright to contribute to the visibilities during the flare.

Assuming ALMA traces particle acceleration \citep{MacGregor:2020} and HETG traces the resulting plasma heating, we only expect ALMA peaks to be associated with the initial period of brightening in the SXR. The initial peak seen in the X-ray occurred during an ALMA calibration gap, but if ALMA had been observing it would likely have recorded a large millimeter peak here too assuming the short timescale flux enhancements arise from particle acceleration during the flare. The first and smaller of the two ALMA peaks reached 18$\pm$4 mJy and the second peak reached 38$\pm$5 mJy, both significantly smaller than previous millimeter-wave peaks \citep{MacGregor:2018, MacGregor:2021}. Only the 38 mJy event lasts long enough to produce a light curve. The ALMA peak appears to occur just after the SXR peak, although this may be an effect of low SXR counts.

The luminosity values corresponding to the 38 mJy and 18 mJy flare are $0.81\pm0.09 \times10^{14}$~erg~s$^{-1}$~Hz$^{-1}$ and $0.36\pm0.08 \times10^{14}$~erg~s$^{-1}$~Hz$^{-1}$, respectively. The spectral index $\alpha$ is defined as $F_\nu \propto \nu^\alpha$ and describes the frequency dependence of the millimeter emission within the band. To obtain $\alpha$, we fit separate point-source models to the visibilities in the lower (213.5 and 216 GHz spectral windows) and upper (228.5 and 230.5 GHz spectral windows) sidebands. The small frequency separation between the lower and upper sidebands provide a weak constraint on the spectral index, leading to large uncertainties for small events. The resulting flux densities of the smaller 18 mJy flare are 17.5$\pm$6.4 and 18.4$\pm$6.6 mJy for the lower and upper sidebands, respectively. Those of the larger 38 mJy flare are 31.5$\pm$6.4 and 43.8$\pm$6.6 mJy, respectively. There is no significant difference between the sidebands for the small peak ($\alpha$=-0.78$\pm$7.9). The large peak is better constrained with $\alpha$=-5.1$\pm$3.9. A lower limit on the linear polarization fraction may be derived from the dual polarization of the observations using Stokes Q and I. The linear polarization fraction is defined as $p_\mathrm{QU}^2 = (Q/I)^2 + (U/I)^2$. The polarization of the smaller flare is undetected while the second and larger flare has a $|Q/I|$=0.18$\pm$0.11.

\subsection{Optical Observations with LCOGT and du Pont}\label{sec:lcogt_obs}
Optical $U$-band photometry was obtained by the Las Cumbres Observatory Global Telescope (LCOGT; \citealt{LCO_Brown:2013}) with a 1 m telescope and Sinistro camera. LCOGT is a suite of 25 telescopes which all work together as a single instrument. Images were obtained at 1.5 min cadence, then dark-subtracted and flat-fielded. Aperture photometry of Proxima and several reference stars was performed to make the light curve. Systematic offsets in the quiescent luminosity near dawn due to the multi-telescope configuration were removed and checked to ensure the flare light curve was not altered. The light curve was converted to fractional flux, $\Delta$F/F=(F-F$_0$)/F$_0$, and the equivalent duration (ED) of the flare was measured in seconds. The $U$-band quiescent luminosity $Q_0$=10$^{26.98}$ erg sec$^{-1}$ is adopted from \citet{Walker:1981} and confirmed using the zero mag flux density, the stellar distance, and Proxima's $U$ mag of 14.21 from \citet{Jao:2014}. The flare energy E$_U$ is computed as ED$\times Q_0$. We find a peak flux of 10$^{28.0}$ erg s$^{-1}$ and integrated energy of 10$^{28.9\pm0.1}$ erg. Similar or larger flares occur in $U$ every 3.3 hr \citep{Walker:1981}.

Optical spectroscopy was obtained the night of the flare at a 600-900 s cadence using the Echelle Spectrograph on the 2.5 m Ir\'en\'e du Pont Telescope at Las Campanas Observatory. The spectrograph operates at a resolution of $R$=40,000 for a 1" slit and provides a complete wavelength coverage of 3500-9850$\AA$. Intermittent clouds resulted in varying S/N over the course of the observations, and several breaks in observing between 5:20 and 6:50 UT. However, we were able to take 21 science exposures encompassing the flare from 3:29 to 8:42 UT. We also took ThAr lamp exposures for wavelength calibration at the beginning and end of the night. 

Spectra were extracted, overscan subtracted, flat-fielded, and wavelength calibrated in \texttt{IRAF} \citep{Tody:1986, Tody:1993}. For each emission line of interest (i.e. H$\alpha$ through H$\gamma$, Ca II H and K, and He I), the continuum in the vicinity of the line was fit with a third order polynomial and subtracted. Equivalent widths for each line were computed by direct integration, where the integration limits were determined by eye and fixed for each line independently. The uncertainty in the equivalent width was estimated from the noise in the continuum on either side of the line. For each spectrum, the limits of integration of the lines are determined by eye, allowing the uncertainties in the continuum to either side of each line to be propagated to the EW values. 

\begin{figure*}
	\centering
	{
		\includegraphics[width=0.85\textwidth]{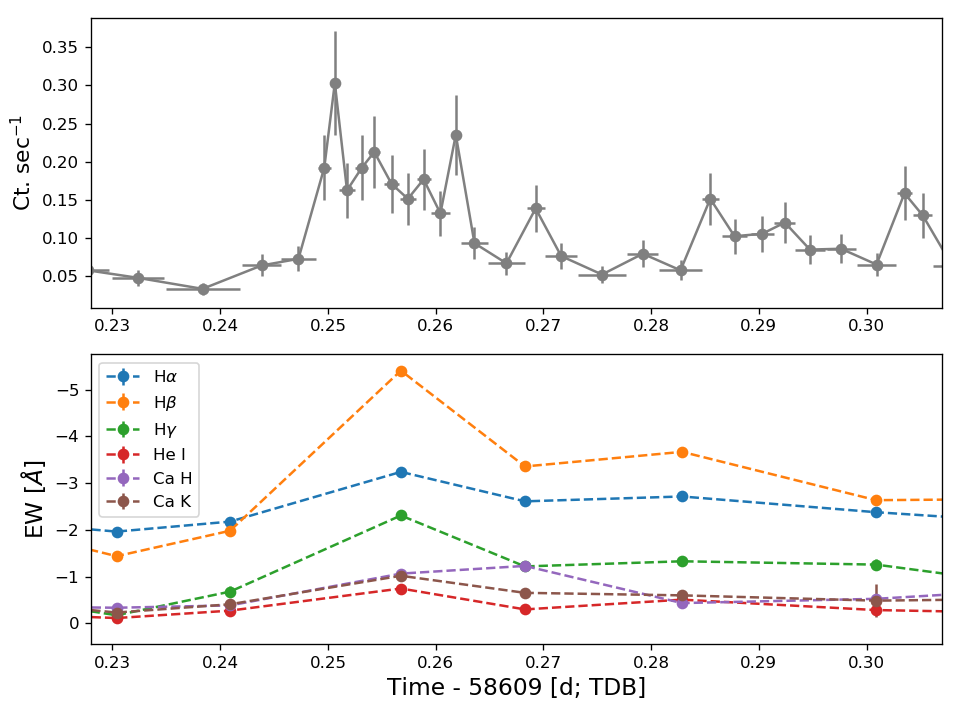}
	}
	\vspace{-0.3cm}
	\caption{Equivalent widths for Du Pont optical line emission on the same time axis as the \textit{Chandra} flare emission but at $\sim$10 min observing cadence. The \textit{Chandra} light curve is shown for reference. Each line but Ca II H appears to peak at the same time as the SXR at this cadence. The EW formal uncertainties are insignificant.}
	\label{fig:dupont_lines}
\end{figure*}

The optical line emission peaks occur at the same time as the SXR peak, at least within the $\sim$10 min Du Pont cadence (Fig. \ref{fig:dupont_lines}). The optical line emission of our flare is weaker than that observed earlier with Du Pont for the extreme millimeter and far UV flare reported in \citep{MacGregor:2021}. The presence of typical flare emission lines in the optical for such a small event supports the argument that millimeter flares are a standard component of the flaring process. 

\begin{figure*}
	\centering
	{
		\includegraphics[width=0.98\textwidth]{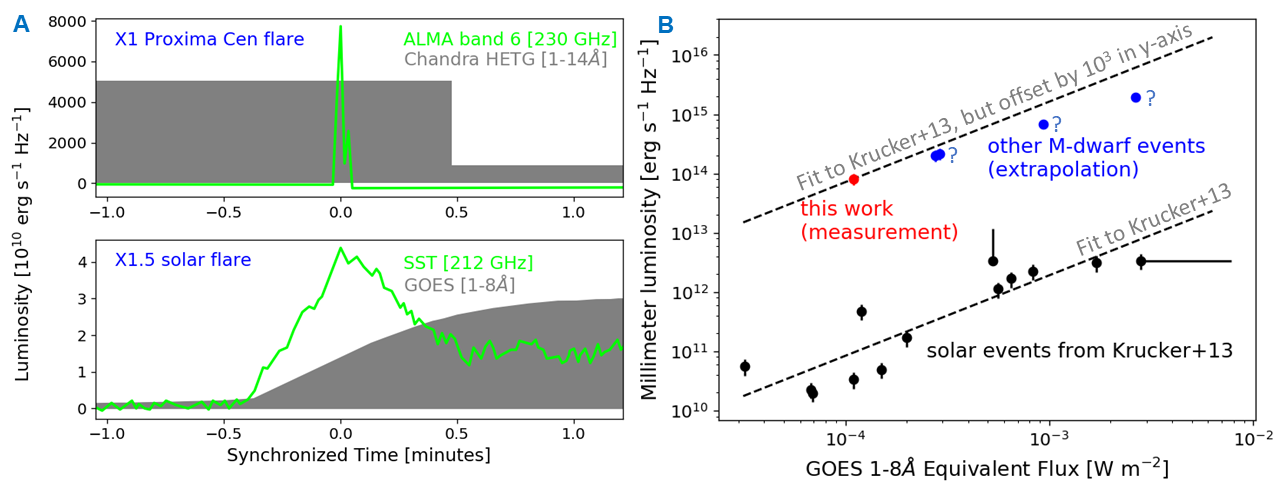}
	}
	\caption{Panel A: the millimeter emission from our X1 flare analogue is both much stronger in intensity and shorter in duration than a X1.5 solar flare (reproduced from \citet{Krucker:2013}). Note the change in the y-axis scaling to illustrate both flares. We only show the second and larger ALMA flare peak as a representative event. Panel B: We reproduce the SXR-millimeter scaling relationship from \citet{Krucker:2013} and overlay our flare in red. We extrapolate the likely position in SXR-millimeter parameter space of larger flares previously observed from M-dwarfs with ALMA by assuming the ratio of SXR and millimeter emission from our flare.}
	\label{fig:vs_K13_flares}
\end{figure*}

\section{Multi-wavelength comparisons to solar and stellar flares}\label{results}
Coordinated X-ray, optical, and millimeter observations of the May 6 event provide an opportunity to explore connections between solar flares and millimeter stellar flares.

\subsection{Overview of correlation of wavelengths in the May 6 Flare}\label{multiwave_descrip}
The millimeter is thought to trace the initial acceleration of charged particles during the impulsive phase of the flare, while optical and SXR wavelengths trace the resulting heating of the stellar photosphere and corona, respectively. Each wavelength observed in the May 6 flare is broadly consistent with this picture. The two peaks of ALMA emission occur on timescales of seconds, consistent with an isolated episode of relativistic particle acceleration. Each peak is accompanied by an increase in $U$-band flux as shown in Fig. \ref{fig:chandra_overview}, consistent with the well-known correlation of $U$-band emission with the impulsive phase of the flare \citep{Gudel:2004a}. While the initial impulsive event was missed by ALMA during its calibrations, the 0.1 mag increase in $U$ traces this initial event during the rapid rise phase of the SXR flare. Each of the other two ALMA events occurs in proximity to a possible corresponding spike in $U$ and the X-ray, although the smaller 18 mJy event is only marginally observed in the other bands. Finally, the SXR heating of the coronal plasma rises rapidly in conjunction with the $U$-band (and ALMA) emission and decays slowly as heat is dispersed.

\subsection{Correlations between Millimeter and SXR emission}\label{SXRwithALMA}
Previous radio observations of stellar gyrosynchrotron emission and likely gyroresonance emission have been obtained at 10-20 GHz frequencies \citep{Gudel:2002}. Gyrosynchrotron emission is optically thick at 10-20 GHz frequencies but optically thin at millimeter frequencies, enabling constraints to be placed on the accelerated electron environment from the spectral index $\alpha$ as described in \citet{Dulk:1985} under the assumption of gyrosynchrotron emission \citep{MacGregor:2020}. Further observations of stellar flares that are obtained simultaneously at 10-20 GHz radio and 230 GHz millimeter frequencies are needed to confirm the identification of ALMA flare events with gyrosynchrotron emission.

While stellar flares have not been observed simultaneously in the SXR and millimeter before, they have been for the Sun. Simultaneous SXR and 210 GHz observations of solar flares ranging in size from M3.2 to $>$X28 class have been reported \citep{Krucker:2013}. As a population, these solar flares show an increase in millimeter emission with increasing SXR flux (Fig. \ref{fig:vs_K13_flares}). If a similar relationship holds for ALMA events, we can extrapolate their SXR emission. If millimeter emission levels from both the Sun and Proxima Cen each trace particle acceleration and SXR emission captures the resulting heating, a correlation for different size flares similar to that observed from the Sun would be a reasonable but unproven assumption. Stellar flares observed by ALMA have luminosities $\geq$10$\times$ higher than their solar counterparts \citep{MacGregor:2021}. Due to the detection threshold of ALMA, we cannot exclude the possibility that a much smaller sun-like component of the millimeter emission is present but undetected. The sharp, strong peak we see with ALMA may be unique to M dwarfs, but lower-level extended emission could be present for both solar and M-dwarf flares.

The presence of SXR emission helps to distinguish whether the millimeter emission of stellar flares correlates with SXR emission as observed for millimeter flares from the Sun \citep{Krucker:2013}. If we tentatively assume for exploratory purposes that the ratio between the SXR and 230 GHz ALMA emission of the May 6 flare (red point in Fig. \ref{fig:vs_K13_flares}) holds for other ALMA stellar flares too, then we would expect larger ALMA events (blue points in Fig. \ref{fig:vs_K13_flares}) to have higher SXR fluxes than smaller events. We also plot the SXR and millimeter emission of the \citet{Krucker:2013} solar flares in the right panel of Fig. \ref{fig:vs_K13_flares} as black points. We fit a line in log-log space to these solar flares in the right panel of Figure \ref{fig:vs_K13_flares}, with each solar flare shown in black. If the fit to the solar flares is shifted upward by a factor of 10$^3$, we note it would nearly go through the extrapolated SXR emission of the ALMA stellar events shown in blue. However, the agreement of the blue ALMA points with the solar flare slope is very tentative with only one stellar flare observed in both bands. If confirmed in a larger sample, the offset between the solar and stellar scaling would suggest a more intense accelerated particle environment in M-dwarf flares than in the Sun.

\begin{figure*}
	\centering
	{
		\includegraphics[width=1.0\textwidth]{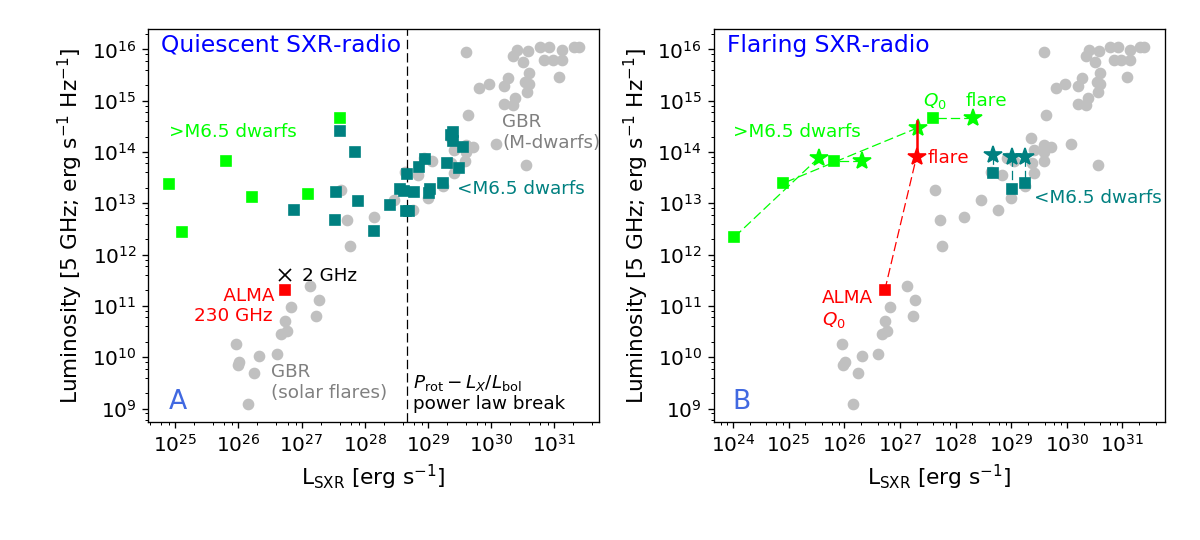}
	}
	\vspace{-0.9cm}
	\caption{The G\"udel-Benz Relation (GBR) describes the relative emission of stars in the SXR and radio, in both quiescent (panel A) and flaring (panel B) states. A departure from the GBR can indicate changes in the emission mechanism; such departures have been observed for radio-loud M-dwarfs with rotation periods $>$10 d \citep{Magaudda:2020} and UCDs, reproduced here from \citet{Williams:2014}. We find the quiescent emission of Proxima Cen to be consistent with the GBR while the flare emission is radio-loud under the assumption of the X1.5 \citet{Krucker:2013} SED. The star's quiescent emission of 0.1 mJy at 230 GHz \citep{MacGregor:2018} is also similar to the $\sim$0.2 mJy observed at 1.6 GHz (black $\times$; \citealt{Perez-Torres:2021}) although this may be electron-cyclotron maser emission. Differences between flares at 5 GHz and 200 GHz are usually within a factor of $\sim$10 during flares, assuming an SED dominated by gyrosynchrotron emission \citep{Krucker:2013}. The X1.5 SED from \citet{Krucker:2013} is used to predict the 5 GHz emission of our flare with an upward error bar. The star symbol is the band 6 value and the error bar is the difference between that and the 5 GHz \citep{Krucker:2013} prediction. Following \citet{Williams:2014}, quiescent emission is shown with colored boxes and flares (colored star shapes) are connected to their quiescent source with dashed lines. $<$M6.5 sources and flares are teal, while $>$M6.5 sources are lime green. The quiescent value of Proxima Cen is from the left-hand panel.}
	\label{fig:vs_radio_flares}
\end{figure*}

Radio-loud late M-dwarfs have been observed to diverge from the G\"udel-Benz Relation (GBR; \citealt{Benz_Gudel:1994}) between the SXR and radio \citep{Williams:2014} as shown in Fig. \ref{fig:vs_radio_flares}. The GBR is a power law relationship between stellar emission in the SXR and radio arising from the underlying processes of magnetic reconnection. Given the unexpectedly high millimeter luminosity seen by ALMA, we explore whether it is consistent with the GBR or the radio-loud spur of \citet{Williams:2014}. Late M-dwarfs inhabit both the GBR and the spur, with stellar rotation appearing to play a role. While too few $<$M6.5 dwarfs in \citet{Williams:2014} lie on the radio-loud spur to be statistically significant without including the $>$M6.5 population, the $<$M6.5 dwarfs appear to begin a transition onto the radio-loud spur near 10$^{28.5}$ erg s$^{-1}$ (Fig. \ref{fig:vs_radio_flares}). We note the radio-loud spur of \citet{Williams:2014} begins at the rotation period of $\sim$10 d and SXR luminosity of 10$^{28.5}$ erg s$^{-1}$ where M-dwarfs transition from saturated to non-saturated emission and the X-ray rotation-activity power law breaks \citep{Magaudda:2020}. Since Proxima Cen rotates at 83 d in the unsaturated regime \citep{Benedict:1998}, it could plausibly lie on the spur. However, Proxima Cen is an M5.5 dwarf and most exceptions to the GBR are $\sim$M7 or later.

While we cannot directly compare ALMA band 6 (230 GHz) millimeter emission with 5 GHz emission, we assume a spectral energy distribution (SED) for an X1.5 flare from \citet{Krucker:2013}. This assumption enables an exploratory comparison of our ALMA flare to be made with the GBR and radio-loud spur, although the comparison is complicated by whether gyrosynchrotron emission \citep{Williams:2014} or coherent emission \citep{Hallinan:2008} is the correct source of the radio-loud behavior. The solar flare SED may also fail to hold in the stellar context given the higher luminosity of the millimeter flares or small sample size of both solar and stellar multi-wavelength flares. Nevertheless, the X1.5 flare SED would predict a 5 GHz luminosity 5$\times$ higher than that seen at 230 GHz in ALMA band 6, which we show as an upper error bar above the band 6 value in Figure \ref{fig:vs_radio_flares}. To determine if the emission appears to be consistent with the GBR, we fit a trendline to the GBR in log-log space and compute the orthogonal distance of each grey point in Fig. \ref{fig:vs_radio_flares} to the GBR in decimal exponents. We find Proxima Cen's quiescent emission in ALMA band 6 is within one standard deviation of the line, and the flare is 5.7 standard deviations above the mean distance in ALMA band 6 and 7.7 deviations assuming the X1.5 solar flare SED. For comparison, the M7 flares on the spur sit 4, 7, 9, and 12 standard deviations from the mean. Also, two quiescent points on the GBR also reach 3-4 standard deviations. Additional uncertainty arises from the comparison of quiescent emission with flare emission, as can be seen in the large offsets of flares from the quiescent emission of the same source in panel B of Fig. \ref{fig:vs_radio_flares}. We find the quiescent emission of Proxima Cen is fully consistent with the GBR while the flare emission is radio-loud. However, the lack of 5 GHz observations and the position of the flare between the GBR and the spur makes it difficult to draw firm conclusions about the flare's place on the GBR.

Millimeter emission of solar flares occurs on timescales of minutes, while that of the May 6 flare and other ALMA stellar flares occurs on timescales of seconds. Millimeter emission in solar flares generally peaks during the rise phase of the SXR flare \citep{Krucker:2013}. In the May 6 flare, the ALMA peaks are associated with two impulsive events in the \textit{Chandra} light curve. It is not possible to tell whether the millimeter emission peaks during the rise phase of the SXR flare as in solar flares, due to the fast nature of the ALMA flare and the low flux of the \textit{Chandra} peak.

The previous stellar flares observed from low-mass stars with ALMA all have negative spectral indices (-2.3$<\alpha<$-1.3), potentially arising from the optically-thin tail of the gyrosynchrotron spectrum. The $\alpha<$-1.2 May 6 flare follows this trend. While most small solar flares (M and $<$X6 class) exhibit falling spectral indices at higher frequencies, many $\geq$X6 solar flares have steeply rising spectral indices resulting from a THz component \citep{Krucker:2013}. It is unclear if the band 6 ALMA observations are seeing the ``normal" falling high-frequency extension of the gyrosynchrotron spectrum or trace other interactions in the flare loop. A complication in identifying the source as optically-thin gyrosynchrotron emission is that increasing the strength of the magnetic field shifts the peak emission to higher frequencies \citep{Krucker:2013}. Given the high magnetic field intensities observed from M-dwarfs \citep{Shulyak:2017}, further modeling work is required. To confirm the emission source in the observational context, wider spectral coverage is required.

\subsection{Correlations between optical and SXR Emission}\label{SXRwithUband}
A relationship between total energy emitted in the optical and in the SXR is well-known from the literature, in which the optical flare energy ranges from $\sim$0.1-1$\times$ the energy emitted in the SXR \citep{Gudel:2004a}. Optical emission in $U$ is well-known to trace the impulsive phase and occurs during the rapid rise of the longer-duration SXR flare. \citet{Kowalski2019} and references therein find that M-dwarf flare spectra peak in the NUV and $U$-band, resulting from the impact of accelerated electrons in the chromosphere. As a result, $U$-band emission peaks during the rise phase of the coronal heating process as described by the Neupert effect \citep{Kowalski:2013}. In order to reproduce observed $U$-band peak flux levels, \citet{Kowalski2019} note models must include both Balmer continuum and blackbody components. As shown in Fig. \ref{fig:chandra_overview} of our flare, the two large $U$-band peaks of 0.1 mag occur during the rise phases of two large impulsive events in the \textit{Chandra} data of the May 6 flare.

The May 6 event is the first stellar flare with both millimeter and $U$-band observations, tying it into the broader context of $U$-band flare observations going back to \citet{Moffett:1974}. Stellar flares are often monitored in $U$-band because the flare spectrum peaks in the region around the near-UV to optical $U$-band. As a result, a number of stellar flares have been observed simultaneously in the SXR and $U$-band \citep{Gudel:2004a}, although the $U$-band energies are not always reported (e.g. \citealt{Schmitt:1993, Gudel:2002, Schmitt:2008}). We compare the relative $U$ and SXR emission during the peak of the flare with literature values using the ratio of the fluxes in erg s$^{-1}$, $L_\mathrm{U}/L_\mathrm{X}$. A small flare from the M6 dwarf UV Ceti observed by \citet{deJager:1989} emitted $L_U$=10$^{29.5}$ erg s$^{-1}$ and $L_\mathrm{SXR}$=10$^{29.1}$ erg s$^{-1}$ at peak, giving $L_\mathrm{U}/L_\mathrm{X}$=2.4. The active early K-dwarf TYC 5315-102-1 emitted $L_U$=10$^{28.2}$ erg s$^{-1}$ and $L_\mathrm{SXR}$=$\sim$10$^{29.9}$ erg s$^{-1}$ at peak, giving $L_\mathrm{U}/L_\mathrm{X}$=0.02 \citep{Pye:2015}. Finally, a flare from the K4/K7.5 dwarf binary By Dra was observed to release $L_U$=10$^{30.2}$ erg s$^{-1}$ and $L_\mathrm{SXR}$=$\sim$10$^{31.9}$ erg s$^{-1}$ at peak, also giving $L_\mathrm{U}/L_\mathrm{X}$=0.02 \citep{deJager:1986}. We note the smaller 0.1-1 range of \citet{Gudel:2004a} holds for total optical energies integrated over longer durations, explaining the greater variability in the $U$-SXR relationship. In this context, our flare's $L_\mathrm{U}/L_\mathrm{X}$=5.0 sits at the top of the literature range. It is notable that the X-ray emission appears suppressed for both the UV Ceti and Proxima Cen flares, both of spectral type M6. Proxima Cen's quiescent $L_\mathrm{U}/L_\mathrm{X}$=1.9 is also high, suggesting a potential spectral type dependence on $L_\mathrm{U}/L_\mathrm{X}$.

Optical line emission such as H$\alpha$ closely follows the SXR (Fig. \ref{fig:dupont_lines}), although the $\sim$10 min cadence of the Du Pont observations makes it difficult to confirm correlations with the other wavelengths. The flare does not show evidence of the delayed H$\alpha$ emission observed in the previous Proxima Cen flare of \citet{MacGregor:2021} or in solar flares \citep{Benz:2017}.

\section{Discussion and Conclusion}\label{discuss_conclude}
We present the first observation of a small stellar flare with \textit{Chandra}, optical, and millimeter coverage. Our flare emitted 10$^{26.5}$ erg in the 1-8 $\AA$ GOES bandpass, which is equivalent to an X1 class flare with a flux of 10$^{-4}$ W m$^{-2}$ at 1 au. The Sun emits 175 similar X1-class flares during its 11 year cycle, providing a unique opportunity to compare the pan-chromatic properties of solar and stellar flares at this energy for the first time.

Like larger millimeter flares, the ratio between millimeter and SXR emission is much higher than observed from the Sun and the ratio of optical to SXR flux is high relative to most solar and stellar flares \citep{Gudel:2004a} but comparable to another M6 dwarf flare from \citet{deJager:1989}. On the other hand, many properties of the flare are surprisingly comparable to its X1 solar counterparts. Its temperature and EM are comparable to M-X solar flares, while being lower than stellar superflares that have temperatures and EMs of 100 MK and 10$^{54}$ cm$^{-3}$, respectively. The relative timing of the SXR, optical and millimeter flare emission is also broadly consistent with flares observed from the Sun. These factors reflect the well-known self-similarity of flare emission properties across a range of orders of magnitude in flare energy, supporting an emerging picture of millimeter emission as a standard component of magnetic reconnection.

The agreement between the quiescent emission in the 230 GHz ALMA \citep{MacGregor:2018} and 1-5 GHz radio band data of the GBR \citep{Benz_Gudel:1994} suggests flux in both bands comes from gyrosynchrotron emission in a population of continually accelerated electrons as suggested by \citet{Williams:2014}. Furthermore, if the gyrosynchrotron SED observed between 210 GHz and 5 GHz for an X1.5 flare in \citet{Krucker:2013} holds for our flare, then it can be compared to the \citet{Williams:2014} radio-SXR flares. On this assumption, the flaring emission appears radio-loud compared to the GBR.

Because it is not clear what causes some M-dwarfs to be radio-loud, our ongoing survey of multiple M-dwarf flare stars with ALMA (2021.1.01209.S) will help to fill out the millimeter/radio-loud and quiet regions of parameter space. Our larger sample will span mid M-dwarfs of various ages and activity levels, enabling us to probe the effect of stellar rotation on the radio-loud spur. It has been suggested that highly-active M-dwarfs have increased radio flare rates at $\sim$1 GHz frequencies relative to the larger population of radio flares from cool stars \citep{Pritchard:2021}. If millimeter flares and 1 GHz radio flares follow a gyrosynchrotron SED, a similar pattern may be observed in ALMA observations of flare stars of various ages and activity levels. Our larger ALMA sample will begin to test this idea. As a larger sample of ALMA flares with SXR spectra are obtained, the magnetic field strength and loop length may also be estimated from the coronal temperature and emission measure of radio-loud flares \citep{Shibata_Yokoyama:2002,Raassen:2007}.

Finally, upcoming observations of Wolf 359 with \textit{Chandra} and ALMA will help to confirm if the high millimeter enhancement for a given SXR flux compared to the Sun is typical or not. If our prediction of an orders of magnitude larger SXR-millimeter scaling relationship for M-dwarf flares compared to solar flares holds, the relationship of particle to thermal emission in M-dwarf flares might also differ. Simulations of 5-100 GHz flux densities resulting from optically-thin gyrosynchrotron emission find a dependence on the high energy electron density and cutoff energy \citep{Wu:2019}. Millimeter emission currently remains one of the few direct probes of the accelerated particle environment in M-dwarf flares \citep{MacGregor:2021}. If millimeter flares result from optically-thin gyrosynchrotron emission during flares, the spectral index $\alpha$ gives the index of accelerated electrons $\delta$ according to the equation $\alpha$=1.22-0.9$\delta$ \citep{Dulk:1985}. By constraining the electron environments of a larger sample of millimeter flares, it may be possible to determine if they are more energetic than those of typical solar flares.

\section{Acknowledgements}\label{sec:acknowledge}
We would like to thank the anonymous referee who graciously gave their time to make this the best version of this work. WH thanks Alexander Brown, Yuta Notsu and Adam Kowalski for helpful conversations on soft X-ray data reduction and optical flares. 

WH acknowledges partial funding support of the Proxima Cen data and analysis through the Cycle 26 HST proposal GO 15651. This material is based upon work supported by the National Aeronautics and Space Administration under award No. 19-ICAR19\_2-0041.

This research has made use of data obtained from the \textit{Chandra} Data Archive and the \textit{Chandra} Source Catalog, and software provided by the \textit{Chandra} X-ray Center (CXC) in the application packages CIAO and Sherpa.

This paper makes use of the following ALMA data: ADS/JAO.ALMA \#2018.1.00470.S. ALMA is a partnership of ESO (representing its member states), NSF (USA) and NINS (Japan), together with NRC (Canada) and NSC and ASIAA (Taiwan), in cooperation with the Republic of Chile. The Joint ALMA Observatory is operated by ESO, AUI/NRAO and NAOJ.

This work makes use of observations from the Las Cumbres Observatory global telescope network. LCOGT data came from proposal NOAO2019A-008.
\software{\texttt{CASA} \cite[v5.3.0 \& v5.11.0][]{McMullin:2007}, \texttt{CIAO} \cite[v4.13][]{Fruscione:2006}, \texttt{Sherpa} \cite[v4.14][]{Freeman:2001, Doe:2007, Burke:2020}, \texttt{XSPEC} \cite[v12.12.0][]{Arnaud1996}, \texttt{IRAF} \citep{Tody:1986,Tody:1993}, \texttt{astropy} \citep{astropy:2013,astropy:2018}}


\bibliography{References.bib}

\end{document}